\documentclass{sig-alternate}

\usepackage{etoolbox}
\patchcmd{\maketitle}{\@copyrightspace}{}{}{}

\usepackage{graphicx}
\usepackage{amsmath,amssymb}
\usepackage[tight]{subfigure}

\usepackage{listings }
\usepackage{algorithm}
\usepackage[noend]{algpseudocode}
\usepackage{hyperref}

\begin{document}

\title{Cheaters in the Steam Community Gaming Social Network}

  \numberofauthors{1}
   \author{
   \alignauthor
        Jeremy Blackburn*, Ramanuja Simha$^{\mp}$, Nicolas Kourtellis*, Xiang Zuo*, \\Clayton Long*, Matei Ripeanu$^{\ddagger}$, John Skvoretz$^{\pm}$, Adriana Iamnitchi* \\
   	\affaddr{*Department of Comp. Science \& Engineering, University of South Florida, Tampa, FL, USA}\\
	\affaddr{$^{\pm}$ Department of Sociology, University of South Florida, Tampa, FL, USA}\\
   	\affaddr{$^{\mp}$ Department of Elec. \& Comp. Engineering, University of Delaware, Newark, DE, USA}\\
	\affaddr{$^{\ddagger}$ Department of Elec. \& Comp. Engineering, University of British Columbia, Vancouver, BC, Canada}\\
	\email{\{jhblackb, nkourtel, xiangzuo, cslong2\}@mail.usf.edu, rsimha@cis.udel.edu, matei@ece.ubc.ca, jskvoretz@usf.edu, anda@cse.usf.edu}
   }
  \maketitle
  
\begin{abstract}
  
  Online gaming is a multi-billion dollar industry that entertains a large, global population.
  One unfortunate phenomenon, however, poisons the competition and the fun: cheating.
  The costs of cheating span from industry-supported expenditures to detect and limit cheating, to victims' monetary losses due to cyber crime.

  This paper studies cheaters in the Steam Community, an online social network built on top of the world's dominant digital game delivery platform. 
  We collected information about more than 12 million gamers connected in a global social network, of which more than 700 thousand have their profiles flagged as cheaters.
  We also collected in-game interaction data of over 10 thousand players from a popular multiplayer gaming server.
  We show that cheaters are well embedded in the social and
  interaction networks: their network position is largely undistinguishable from that of fair players.
We observe that the cheating behavior appears to spread through a
social mechanism: the presence and the number of cheater friends of a
fair player is correlated with the likelihood of her becoming a
cheater in the future. 
Also, we observe that there is a social penalty involved with being
labeled as a cheater:  cheaters are likely to switch to more
restrictive privacy settings once they are tagged and they
lose more friends than fair players.
Finally, we observe that the number of cheaters is not correlated with the geographical, real-world population density, or with the local popularity of the Steam Community.

This analysis can ultimately inform the design of mechanisms to deal
with anti-social behavior (e.g., spamming, automated collection
of data) in generic online social networks.

\end{abstract} 
  %
  %
  %

\section{Introduction}\label{intro}

The popularity of online gaming led to the creation of a billion dollar industry, but also to a vigorous cheat code development community that facilitates unethical in-game behavior. 
``Cheats'' are software components that implement game rule violations, such as seeing through walls or automatically targeting a moving character. 
It has been recently estimated that cheat code developers generate between $\$15,000$ and $\$50,000$ per month from one class of cheats for a particular game alone~\cite{apb_blog}.

In all cultures, players resent the unethical behavior that breaks the rules of the game: ``The rules of a game are absolutely binding [...] As soon as the rules are transgressed, the whole play-world collapses. The game is over~\cite{huizinga50homo-ludens}''.   
Online gamers are no different judging by anecdotal evidence, vitriolic comments against cheaters on gaming blogs, and the resources invested by game developers to contain and punish cheating (typically through play restrictions). 
For some cheaters, the motivation is monetary. 
Virtual goods are worth real-world money on eBay, and online game economies provide a lucrative opportunity for cyber  criminals~\cite{Keegan:2010hg,Ku2007online-gaming-crime}. 
For other cheaters, a competitive advantage and the desire to win is motivation enough~\cite{Nazir:2010p910}.

Cheating is seen by the game development and distribution industry as both a monetary and a public relations problem~\cite{consalvo07-cheatingbook} and, consequently, significant resources are invested to contain it.
For example, \emph{Steam}, the largest digital distribution channel for PC games, employs the Valve Anti-Cheat System (VAC) that detects cheats and marks the corresponding user's profile with a permanent, publicly visible, red, ``ban(s) on record''. 
Game servers can be configured to be VAC-secured and reject players with a VAC-ban on record matching the family of games that the server supports. 
The overwhelming majority of servers available in the Steam server browser as of October 2011 are VAC-secured. 
For example, out of the 4,234 Team Fortress 2 servers available on October 12, 2011, 4,200 were VAC-secured.
Of the 34 non-secured servers, 26 were servers owned and administrated by a competitive gaming league that operates its own anti-cheat system.

Gaming mimics, to some extent, real-world interactions~\cite{Szell:2010p1131}. 
Understanding the cheaters' position in the social network that connects gamers is relevant not only for evaluating and reasoning about anti-cheat measures in gaming environments, but also for studying social networks at large. 
Cheaters are unethical individuals who can model the position of individuals in large-scale non-hierarchical communities that abuse the shared social space. 
In online social networks, they can model the abuse of available, legal tools, such as intensive use of communication tools for political activism. 
Taken to the extreme, such behavior leads to the tragedy of the commons: all players become cheaters and then abandon the game; corruption escalates and chaos takes place; and communication is buried in noise. 

Like many gaming environments, Steam allows its members to declare social relationships and connect themselves to \emph{Steam Community}, an online social  network. 
This work reports on our analysis of the Steam Community social graph with a particular focus on the position of the cheaters in the network.
To enable this study, we crawled the Steam Community and collected data for more than 12 million users. 
Our analysis targets the position of cheaters in the networks; evidence of homophily between cheaters; geo-social characteristics that might differentiate cheaters from the fair population; and the social consequences of the publicly visible cheating flag. 

 Our study shows that cheaters are well embedded in the social network; they exhibit a high degree of homophily; their geo-social characteristics differ from those of non-cheaters; and while the cheating flag does not a affect their aggregate well being in the gaming environment, it is penalized by friendship loss, shorter in-game interactions, and marked by some degree of embarrassment (as suggested by more frequent changes to private profiles).  Additionally, our temporal analysis of the cheating data suggests that cheating behavior spreads via social relationships: the presence and the number of cheater friends of a fair player is correlated with the likelihood of her becoming a cheater in the future.

An overview of related work is presented in Section~\ref{related}. 
The datasets we collected are presented in Section~\ref{sec:datasets}, along with our data collection methodology.
Section~\ref{sec:cheaters-friends} analyzes the position of cheaters in the network from the perspective of declared relationship, in-game interactions, and the strength of their relationships measured via social-geographical metrics. 
It also presents the effect of the VAC-ban on individual players.
Section~\ref{propagation} reasons about possible mechanisms for spreading the cheating behavior. 
Section~\ref{sec:discussions} concludes with a summary of our findings and their consequences.

\section{Related Work}
\label{related}



\emph{Cheating in social gaming} is a relatively unexplored area.
Nazir et al. study fake profiles created
to gain an advantage in social gaming contexts
in~\cite{Nazir:2010p910}. Through the evaluation of behavior of player
accounts within Fighters' Club (FC), a game on the Facebook Developer
Platform, they are able to predict with high accuracy whether a
profile is fake.
Users in FC cheat by creating fake profiles to gain an advantage (i.e., they perform a Sybil attack), whereas cheaters in Steam Community are not trying to alter the structure of the social graph.
Instead, they are attacking game rule implementations.

``Gold farmers'' are cheaters that make black market exchanges of real world currency for virtual goods outside of sanctioned, in game, trade mechanisms.
By examining social networks constructed from database dumps of banned \emph{EverQuest II} (EQ2) players, Keegan et al. found gold farmers exhibit different connectivity and assortativity than both their intermediaries and normal players, and are similar to real-world drug trafficking networks~\cite{Keegan:2010hg}.
Ahmad et al.~\cite{Ahmad:2011wp} further examined trade networks of gold farmers and the items they trade and propose models for deviant behavior prediction.

Their data set differs from ours in both motivation for cheating, and the method of punishing cheaters.
No clear financial motivation for cheating exists in the majority of games played by Steam Community players.
Additionally, while cheaters in EQ2 have their accounts permanently disabled, cheaters in Steam Community are only restricted from playing the particular game they were caught cheating in on VAC-secured servers, as explained in Section~\ref{sec:datasets}.

Finally, we note that to the best of our knowledge, this is the largest scale study of cheaters in a gaming social network.
We discovered over double the amount of cheaters as there were players in~\cite{Nazir:2010p910}, and multiple orders of magnitude more cheaters than players in~\cite{Keegan:2010hg,Ahmad:2011wp}.

Although not much quantitative analysis has been performed, cheating in video games has been studied qualitatively.
Duh and Chen describe several frameworks for analyzing cheating, as well as how different cheats can impact online communities in~\cite{Duh:2009vk}.
\emph{Neopets}, a web based social game, was examined by Dumitrica in~\cite{Dumitrica:2011jd}.
She describes a process by which gamers, who naturally seek ways to increase their ``gaming capital'', are tempted to cheat, and argues that a cheating culture emerges from social games, where social values are used to understand and evaluate the ethical questions of cheating.


\emph{Social networks of online gamers} have been addressed in recent
studies. Szell and Thurner~\cite{Szell:2010p1131} provide a detailed analysis of the
social interactions between players in Pardus, a web-based Massively
Multiplayer Online Game (MMOG).
They employee, as we do in this study, traditional tools from social
network analysis, however, there are a few significant differences in
the datasets used.
First, declared relationships between users in the Steam Community OSN are \emph{informed} by underlying in-game interactions, but exist
in a more general ``gaming'' context than in Pardus, where the social network is built
on interactions within the context of one game. Second, players in
Pardus can declare friends and enemies. In our study, players can only declare friends.
While we do provide some results based on interaction data from a Team Fortress 2 (TF2) server, TF2 players do not declare friends and foes, but compete on ad-hoc, opposing teams.
Finally, while there might be cheaters present in the Pardus dataset, they are not identified or studied in anyway.

Xu et al.~\cite{Xu:2011tu} interviewed 14 Halo 3 players to study the meaning of relationships within an online gaming context.
Halo 3 is a multiplayer First Person Shooter (FPS) available on the Xbox game console, similar in style and stature to the most played games on \emph{Steam}.
They found evidence of in-game relationships being supported by real-world relationships, triadic closure of relationships making use of both real and virtual relationships as a bridge, and in-game interactions strengthening ties in the real world.
They further found evidence of social control as a tool for managing deviant behavior.
In addition to in game interactions on a single game server, we also measure and analyze the social structure of millions of online gamers and their relationships with the deviant class of users that cheat.

\emph{General gaming studies} have been prompted by the popularity of online gaming.
There has been significant interest in
understanding the technological needs for supporting gaming
platforms. Consequently, various studies characterized network traffic
due to gaming, resource provisioning, work load prediction,
and player churn in online
games~\cite{Chambers:2010wb,Claypool:2003dh,Feng:2002dx,WuchangFeng:ce,Feng:2003iu}. Other
studies have  focused on the psychological and social properties of
gamers~\cite{Fritsch:2006:DOH:1230040.1230082} and gaming  communities~\cite{Balint:2011ua,Wei:2010p911,Ducheneaut:2007if}.

\section{Datasets}\label{sec:datasets}

In order to better understand the datasets, we start by describing the Steam Community network. 
Run by Valve, who also develops some of the most successful multiplayer first-person shooter games, Steam controls between 50\% and 70\% of the PC digital download market and is more profitable per employee than Google and Apple~\cite{steam_pcgamer}. 
It claims more that 30 million user accounts as of October 2011.

While games from a number of developers and publishers are available
for purchase on Steam, an important segment is formed by the \emph{multiplayer} FPS genre. 
In contrast to massively multiplayer online games, multiplayer FPSs usually take place in a relatively ``small'' environment, player actions generally do not affect the environment between sessions, and instead of one logical game world under the control of a single entity, there are multiple individually-owned and operated servers.
  Because there is no central entity controlling game play and since there is a very large number of servers to choose from, 
 the communities that form around individual servers are essential to the prolonged health of a particular game.  

\subsection{The Steam Community}\label{sec:community}

Recognizing the social nature of
gaming in general, Valve created the \emph{Steam Community}. 
Steam Community is a social network comprised of Steam users, i.e., people who buy and play games on
Steam.
To have a Steam Community profile, one first needs to have a Steam account and take the additional step of configuring a profile.
Users with a Steam account and no profile (and thus, not part of the Steam Community) can participate in all gaming activities, and can be befriended by other Steam users, but no direct information is available about them. 
Steam profiles are accessible in game via the Steam client and are also available in a traditional web based format at \url{http://steamcommunity.com}.

Valve also provides the Valve Anti-Cheat (VAC)  service that detects players who cheat and marks their profiles with a publicly visible, permanent \emph{VAC ban}.
Server operators can ``VAC secure'' their servers: any player with a  VAC ban for a given game can not play that game on VAC secured servers (but they are allowed to play other games).
In an effort to stymie the creators and distributors of cheats and hacks, the details of how VAC works are not made public. 
What is known is that VAC bans are not issued immediately upon cheat detection, but rather in delayed waves, as an additional attempt to slow an arms race between cheat creation and detection.

While Steam accounts are free to create, they are severely restricted until associated with a verifiable identity, such as resulted from game purchases (via a credit card) or from a gift from a verified account.
Once associated with an account, game licenses (whether bought or received as a gift) are non-transferable.
This serves as a disincentive for users to abandon flagged accounts for new ones: abandoning an account means abandoning all game licenses associated with that account.
Moreover, Sybil attacks become infeasible, as they would require monetary investments and/or a real-world identity even for the most trivial actions, such as chatting with other players.

\subsection{Data Collection}\label{crawler}

In our analysis we used three data sources. 
The vast majority of our data was obtained by crawling the Steam Community website to collect user profiles and the resulting social network. 
In order to augment profile information with the (approximate) time of VAC bans, we queried the \url{vacbanned.com} site. 
And finally, we obtained in-game interactions from a \emph{Team Fortress 2} server located in California.

\textbf{Crawling the Steam Community:} 
At the time of the data collection, a rate-limited web API for accessing Steam was available, but it was restricted to summary information that excluded friend lists. 
Unfortunately, in addition to being limited to 100,000 calls per day, the web API does not provide access to the friend list, but only to limited summary information for a user.
As an alternative, Steam Community data is made available via unmetered, consumable XML. 
Using the unmetered, consumable XML on the Steam Community web site, we crawled during March 16th and April 3rd, 2011.
The majority of the data (over 75\%) was collected between March 25th and March 31st.
The crawler collected user profiles by starting from a randomly generated set of SteamIDs and following the friendship relationships declared in user profiles. 
To seed our crawler, we generated 100,000 random SteamIDs within the key space (64-bit identifiers with a common prefix that reduced the ID space to less than $10^9$ possible IDs), of which 6,445 matched configured profiles. 

The crawling was executed via a distributed breadth first search.
Each of the initial seed SteamIDs was pushed onto an Amazon Simple Queue Service (SQS) queue. 
Each crawler process  popped one SteamID off this queue and retrieved the corresponding profile data via a modified version of the Steam Condenser library.
The profile data of the crawled user was stored in a database and any newly discovered users (i.e., friends that were previously unseen) were added to the SQS queue.
Crawling proceeded until there were no items remaining in the queue. 
Using RightScale, Inc's cloud computing management platform to automatically scale the crawl according to the number of items in the SQS queue, we ended up running up to six Amazon ``c1.medium'' EC2 instances executing up to 15 crawler processes each. 

A Steam profile includes a nickname, a privacy setting (public, private, friends only or in-game only), set of friends (identified by \emph{SteamIDs}), group memberships, list of games owned, gameplay statistics for the past two weeks, a user-selected geographical location, and a flag (VAC-ban) that indicates whether the corresponding user has been algorithmically found cheating. 
We augmented the information for the VAC-banned players with a timestamp that signifies when the VAC ban was first observed (as explained next).

From our initial $6,445$ seeds of user ids, we discovered just about $12.5$ million user accounts, of which 10.2 million had a profile configured (about 9 million public,  313 thousand private, and 852 thousand visible to friends only). 
There are 88.5 million undirected \emph{friendship} edges and  1.5 million user-created groups. 
Of the users with public profiles, 4.7 million had a location set (chosen from 33,333 pre-defined locations),  3.2 million users with public profiles played at least one game in the two weeks prior to our crawl, and 720 thousand users are flagged as cheaters.
Table~\ref{tbl:dataset} gives the exact numbers.

\begin{table*}[t]
  \centering
  \begin{tabular}{| r | r | r | r | r | r | r | r | r |}
    \hline
    \multicolumn{1}{|c|}{Account} & \multicolumn{1}{|c|}{$All$} & \multicolumn{1}{|c|}{$Edges$} & \multicolumn{1}{|c|}{$Profiles$} & \multicolumn{1}{|c|}{$Public$} & \multicolumn{1}{|c|}{$Private$} & \multicolumn{1}{|c|}{$Friends-Only$} & \multicolumn{1}{|c|}{$Location-Set$}\\
    \hline
    All users & 12,479,765 & 88,557,725 & 10,191,296 & 9,025,656 & 313,710 & 851,930 & 4,681,829\\
    \hline
    Cheaters & - & - & 720,469 & 628,025 & 46,270 & 46,174 & 312,354\\
    \hline    
  \end{tabular}
  \caption{
Size of the \emph{Steam Community} dataset.
  }
  \label{tbl:dataset}
\end{table*}

\begin{figure}[htbp]
	\includegraphics[width=3in]{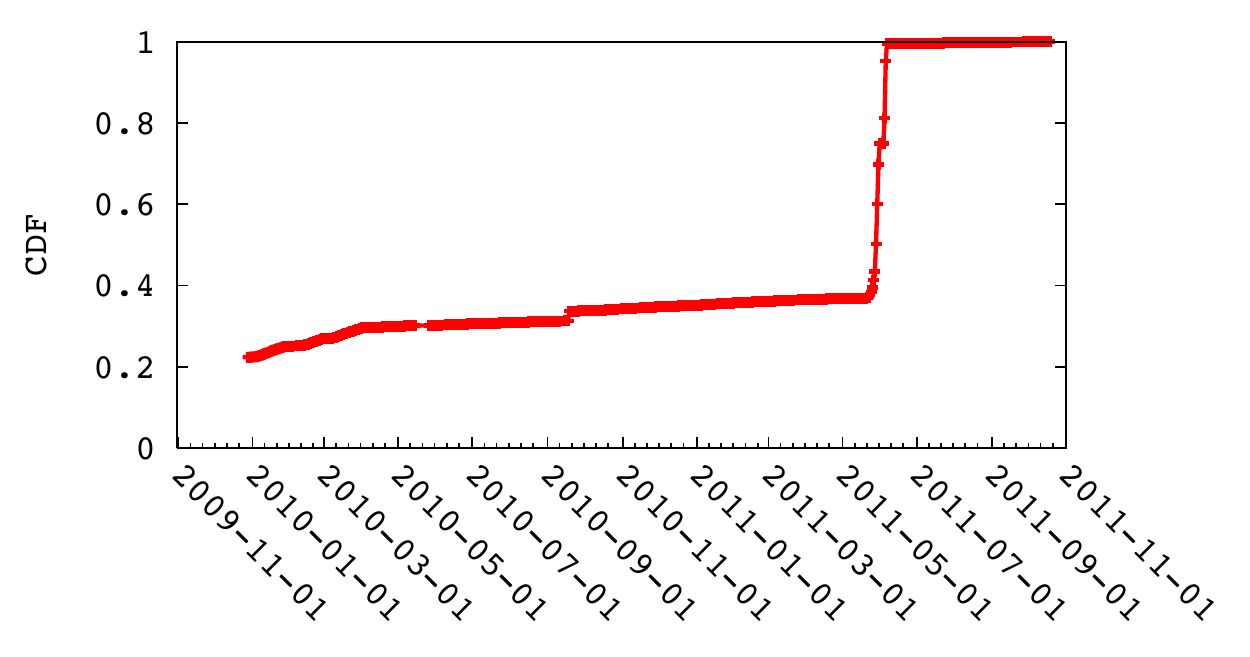}
	\caption{Historical VAC ban dates as reported by vacbanned.com.
  	The date of discovery is on the x-axis and the number of new banned accounts discovered is on the logscale y-axis.
  	The jump around end of May 2011 is probably due to an effort from the website to populate their database.
  	}
\label{fig:banned_since_distribution}
\end{figure}

\textbf{Collecting VAC Ban Timestamps: } We collected historical data on when a cheating flag was first observed from a 3rd party service, \url{vacbanned.com}, that allows users to enter a SteamID into a search box to check whether or not that SteamID has been banned.
If the account is banned, the date the ban was first observed is provided.
We also re-crawled (between Oct. 18th  and Oct. 29th 2011) all Steam profiles discovered during the first crawl without a VAC ban, to identify which non-cheaters had been flagged as cheaters since April 2011.
Of these, $43,465$ now have a VAC ban on record.

\url{Vacbanned.com} had observed ban dates for 423,592 of the cheaters we discovered during our initial crawl.
Figure~\ref{fig:banned_since_distribution} shows a CDF of these ban observations over time.
The earliest dates indicate users that were banned prior to December 29th, 2009.
We combined the ``banned-since'' dates from our original crawl, \url{vacbanned.com}, and our re-crawl.
In the case of a user profile having more than one ban date (due to the 3 sources),
the earliest date was chosen.
It is important to note that all ban dates were treated as ``on or before'' as opposed to a precise timestamp.
This is because
the ban dates are when the ban was first observed by a 3rd party (\url{vacbanned.com} or our crawler), not necessarily when it was applied by Valve.

\textbf{In-game interactions:} We have acquired detailed game play logs of a 32-simultaneous player VAC-secured Team Fortress 2 (TF2) server located in California.
TF2 is a critically-acclaimed, team-based, objective-oriented first-person shooter game and is played on thousands of servers at any given time.
Our logs span just over 2 months from April 1 to June 8, 2011, and consist of various game-specific events involving $10,354$ players.
Because this server is VAC-secured, no players that have cheated in TF2 appear in the logs; the only cheaters that appear are those that were caught in a different game.

From the server logs we extracted 5 different interactions types between users, and constructed an interaction graph where an edge exists between two players if they interacted together during the game.
The resulting graph contained $10,354$ players of which $93$ were cheaters and had $486,808$ edges.

\section{Cheaters and their gaming habits}

Although the majority of this work is concerned with how cheaters and non-cheaters are positioned within the Steam Community, it is worthwhile to first examine differences in their behavior as gamers.
This section explores and identifies how cheaters and non-cheaters differ with respect to socio-gaming properties as well as game play and purchasing habits.

\subsection{Are cheaters social gamers?}

Although previous work has indicated that gaming in general is a social activity, one might question whether cheaters, easily considered as anti-social actors by their very nature, also engage in gaming as a social activity.
As gaming is a social activity we should expect users with more friends to  correspondingly invest more in the opportunity to game with those friends.
E.g., more games owned (to widen the audience of potential play partners) and more hours played (to increase the interaction time with said play partners).
     
     \begin{figure}[th]
     \centering
     \subfigure[Games owned as a function of degree.]{
     \includegraphics[scale=0.9]{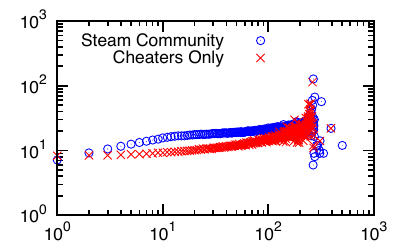}
     \label{fig:degree_gamesowned}
     }
     \subfigure[Hours played as a function of degree.]{
     \includegraphics[scale=0.9]{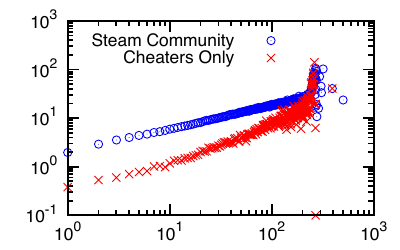}
     \label{fig:degree_hoursplayed}
     }
     \caption{The number of games owned, and hours played in the past two weeks as function of degree.}
     \label{fig:deg_vs_game_stats}
     \end{figure}

Figures~\ref{fig:degree_gamesowned} and~\ref{fig:degree_hoursplayed} plot the average number of games owned and hours played in the two weeks prior to our crawl as a function of degree, respectively.
From Figure~\ref{fig:degree_gamesowned}, we observe that after a quick rise in the number of games owned up to about 10 friends, there seems to be a slightly increasing relationship between the number of friends a user has and the number of games she owns, and that for cheaters, even though they own fewer games than non-cheaters on average, this trend is more visible.
We see a similar pattern in Figure~\ref{fig:degree_hoursplayed}, with slightly more acute response from the cheaters.

We suspect these increasing trends have to do with ``peer marketing.''
Simply put, users see their friends playing games, and make both purchasing and playtime decisions based on this.
As many games have such a heavy multiplayer focus, the more friends a user has the more likely one of those friends will be available to play any given game.
This makes sense when viewing gaming as a social activity: the more friends you have, the more opportunity you have to play.
Further, because friendships on \emph{Steam Community} are likely to form due to in-game relationships and experiences, the more hours a user plays, the greater her chance to create new friendships.
Ultimately, we believe Figure~\ref{fig:deg_vs_game_stats} is indicative of a feedback cycle where users discover new friends from the games they play, and discover new games from the friends they play with.

In other words, the investment in gaming (both monetary and time) increases with the number of friends for both cheaters and non-cheaters.
Even though cheaters are involved in decidedly anti-social behavior, they still have a positive response to the social phenomena of gaming.
This is an important result, as it introduces to the possibility of VAC bans having more than just a utilitarian impact on cheaters: not only is their technical ability to game affected, but the ban might also effect their standing with their gameplay partners.
In fact, in Section~\ref{sec:cheater-changes} we show that there are indeed negative social effects associated with being branded a cheater.

\subsection{What kind of games do cheaters play?}

\begin{figure*}[ht]
  \centering  \includegraphics[scale=1.0]{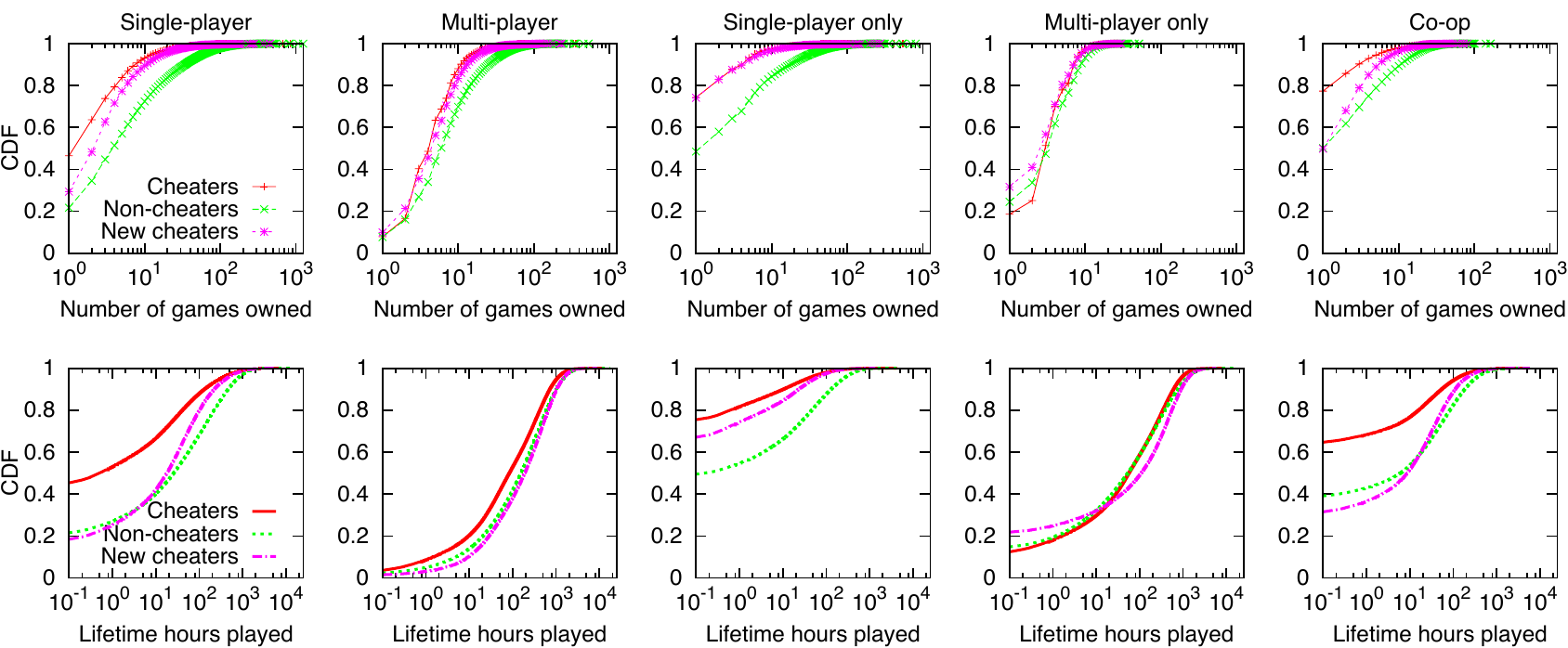}
  \caption{CDF of the number of games owned and lifetime hours per category.}
  \label{fig:games_owned_hours_played_per_category}
\end{figure*}

As just described, cheaters' gaming habits are positively correlated with their social position.
However, this does not address questions regarding the \emph{kind} of gaming cheaters partake in.
There are numerous ways to classify individual games, and the Steam Store tags each games with a variety of categories.
We decided to use the categories ``single-player'' (the game can be played by a single human player) and ``multi-player'' (the game supports multiple human players).
This is a natural categorization for our purposes as VAC bans only have an affect on multi-player games, and all games are tagged in \emph{at least one} of these categories.
Some games do not contain a single-player component at all (e.g., TF2).
We classified these types of games as ``multi-player only'' if they were tagged as multi-player but not single-player, and those with no multi-player component are likewise classified as ``single-player only''.
Finally, we made use of one additional category:``co-op''.
Co-op, or cooperative, games are loosely defined as multi-player games with a mechanic focusing on co-operative (as opposed to antagonistic) interaction between human players.
For example, players might work together to defeat a horde of computer controlled goblins~\cite{dungeon-defenders}, or to excavate a landscape and build a city~\cite{terraria}.

Figure~\ref{fig:games_owned_hours_played_per_category} plots the number of games owned and and the lifetime hours on record per game category for cheaters, non-cheaters, and the newly flagged cheaters discovered in our October re-crawl.
First, we can see further confirmation of gaming as a social-activity: gamers on Steam Community are far more likely to own more than one multi-player games than single-player games, even though there are over twice as many single-player games available on Steam than there are multi-player games.
This trend is even clearer when considering single-player only games vs. multi-player only games.
Next, we observe that non-cheaters are more likely to own more games than cheaters in general .
However, we note that cheaters are slightly more likely to own more than two multi-player only games when compared to non-cheaters, and that the difference in number of games owned between cheaters and non-cheaters is smaller for multi-player games than for single-player games.
This is further indication that cheaters are social gamers: even though they might not own as many games as a whole, they are as interested in multi-player games as non-cheaters are.

When considering the lifetime hours played per category, we see a somewhat different story.
The original cheaters played far fewer hours of single-player games when compared to both the non-cheaters and the newly flagged cheaters.
Further, the newly flagged cheaters and the non-cheaters have very similar CDFs for hours played in single-player, multi-player (only), and co-op games, however, this is not true for single-player only games.
This might be due to the classification of currently popular (for cheating) games.
In any event, we see that cheaters are most definitely social gamers, favoring multi-player games over single-player games for both purchase and play time.
Specifically, cheaters are much less interested in games \emph{without} a multi-player component.

\section{Cheaters and Their Friends}\label{sec:cheaters-friends}

One line of thought in moral philosophy is that (non)ethical behavior of an individual is heavily influenced by his social ties~\cite{parfit1984-reasons}. 
Under this theory, cheaters should appear tightly connected to other cheaters in the social network. 
On the other hand, unlike in crime gangs as the ones presented in~\cite{Ahmad:2011wp}, cheaters do not need to cooperate with each other to perform their actions. 
Moreover, playing against other cheaters may not be particularly productive. 
These observations suggest that cheaters may be dispersed in the network, contradicting thus the first intuition. 

To understand the position of cheaters in the social network, we investigate their relationships as measured by 1)~their number of friends and their cheating status (Section~\ref{degrees}), 2)~the in-game interactions with other players (Section~\ref{in-game}) and 3)~sources of social closeness (Section~\ref{geosocial}).

\subsection{Who is Friends with Cheaters?}\label{degrees}

\begin{figure}[htbp]
        \centering
	\subfigure[All users and cheaters.]{
		\includegraphics[width=1.5in]{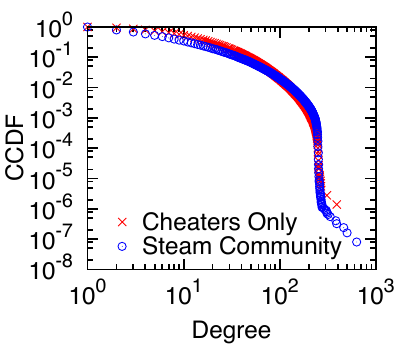}
		\label{fig:deg_dist_steam_comm}
	}
	\subfigure[Public, private, friends only, and no profile.]{
	\includegraphics[width=1.5in]{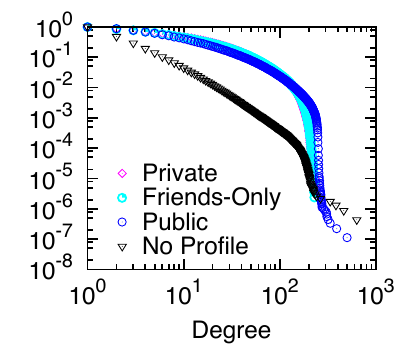}
	\label{fig:deg_dist_inferred}
	}
\caption{Degree distributions for the \emph{Steam Community}}
\label{fig:deg_dist}
\end{figure}

The degree distribution of the \emph{Steam Community} graph as a whole, just cheater profiles, as well as for private, friends-only profiles, and users without profiles are plotted as CCDF in Figure~\ref{fig:deg_dist}.
For users without a profile or private profiles, edges in the graph are inferred based on the information from public profiles that declare the user as a friend.
From the degree distributions we make two observations.

First, we discovered a hard limit of 250 friends. 
However, there are some users who have managed to circumvent this hard limit. 
One user in particular has nearly 400 friends, and through manual examination we observed this user's degree increasing by one or two friends every few days.
Coincidentally, this profile also has a VAC ban on record.   

Second, all categories plotted in Figure~\ref{fig:deg_dist}, with the exception of that of users with Steam accounts but no profiles, overlap, fitting the same power law distribution (coefficient of $-0.92$).  
Consequently, cheaters have about the same number of declared friends as non-cheaters.
The result also shows that attempting to hide connection information through private or friends-only privacy settings is unsuccessful: in this case, the player's privacy is determined by the privacy settings of his friends.  

While cheaters are mostly indistinguishable from non-cheaters using the node degree distribution, a more important question is whether or not cheaters act in complete isolation or if their deviant behavior shows network effects.
In other words, are cheaters more likely to be friends with other cheaters than with non-cheaters?

\begin{figure}[htbp]
\centering
\subfigure[]{\includegraphics[width=1.5in]{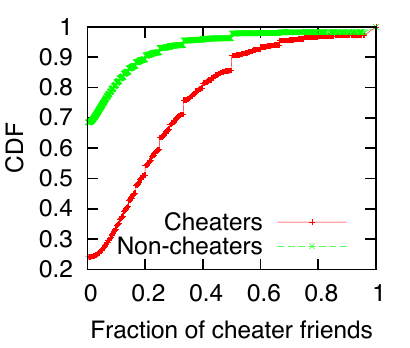}\label{fig:fraction_cheater_friends_cheaters_vs_noncheaters}}
\subfigure[]{
\includegraphics[width=1.5in]{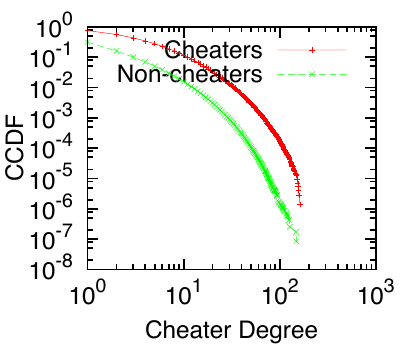}\label{fig:cheating_degree_ccdf}}
\caption{
(a) Fraction of cheaters' friends that are cheaters vs. the fraction of non-cheaters' friends that are cheaters.
(b) CCDF of the number of cheaters' friends that are cheaters vs the number of non-cheaters' friends that are cheaters. I.e., the ``cheating degree'' of the declared friendship network for cheaters and non-cheaters.}
\label{fig:cheater_growth}
\end{figure}

Figure~\ref{fig:fraction_cheater_friends_cheaters_vs_noncheaters} plots the CDF of the fraction of a player's friends who are cheaters.  Figure~\ref{fig:cheating_degree_ccdf} plots the CCDF of the number of cheaters friends for both cheaters and non-cheaters when varying the node degree. (This figure is comparable to Figure~\ref{fig:deg_dist_steam_comm}, but displays only the ``cheating'' degree of users).

The picture that emerges from these two figures is a striking amount of homophily between cheaters: cheaters are more likely to be friends with other cheaters. 
While nearly 70\% of the non-cheaters have \emph{no} friends that are cheaters, 70\% of the cheaters have at least 10\% cheaters as their friends. Roughly 15\% of cheaters have over half of their friends other cheaters.

We next estimated the diameter of the Steam Community using the Hadoop-based DIameter estimator ($HADI$)~\cite{kang11hadi}.
$HADI$ estimates the maximum, average and effective diameter of a graph by computing the neighborhood function $N(h)$, which refers to the number of reachable pairs of nodes within $h$ hops of each other.  
This estimator iteratively increases the number of hops $h$ and stops when the difference $N(h)-N(h-1)$ is zero or the search has stabilized to a set of nodes.  

We report on the following three diameter measures as reported by $HADI$. 
\begin{itemize} 
\item The \emph{maximum} diameter $h_{max}$ is the value of $h$ where the  iteration process stops. 
 
\item The \emph{effective} diameter is the minimum number of hops in which $90\%$ of all connected pairs of nodes can reach each other, i.e., when $N(h)=0.9N(h_{max})$. 

\item The \emph{average} diameter is the expected value of $h$ over the distribution of $N(h)-N(h-1)$, i.e., $\sum_{h=1}^{h_{max}}h * (N(h)-N(h-1))/(N(h_{max})-N(0))$. 
\end{itemize}

Table~\ref{tab:diameters} presents these diameter measures for the Steam Community as a whole, as well as the network composed only of cheater-to-cheater edges (C-C).
Furthermore, the cheaters-only network is about 45\% more stretched in comparison to the full social network, signifying that cheaters must expend a lot more effort to reach each other through only cheater-to-cheater edges.
The effective and average diameters, however, vary much less: a user can reach about 90\% of the population in this planetary-scale network within a maximum of $6$ to $11$ hops, even in the less tightly connected cheaters subgraph.
Previous studies reported similar results, such as the low average path length of $6.6$ hops between users in the world-wide distributed MSN population~\cite{leskovec08msn}.

\begin{table}[ht]
\begin{center}
\begin{tabular}{|c|c|c|c|}
\hline
Network &  \multicolumn{3}{c|}{Diameter} \\
\cline{2-4}

						&	Maximum 	&	Effective 	&	Average 	\\
\hline \hline
Social 		&	15				&	7.01				&	6.22				\\ \hline
Social C-C	&	22				&	8.61				&	7.00				\\ \hline
\end{tabular}
\end{center}
\caption{Diameter measures. The maximum diameter of the cheaters-only
  subgraphs is significantly
  larger than the average and effective diameters.}
\label{tab:diameters}
\end{table}

\begin{figure}[ht]
\begin{center}
	\includegraphics[scale=0.65]{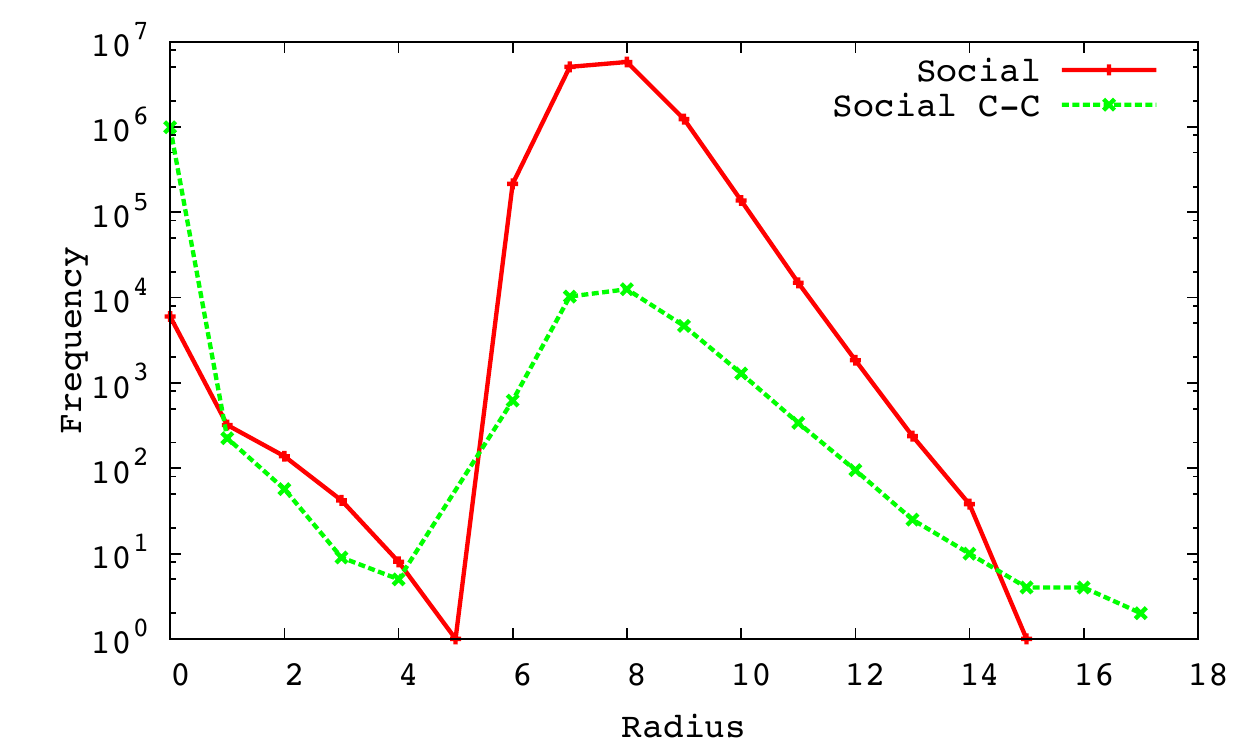}
\caption{Distribution of radius in the Steam Community as a whole and the cheater-to-cheater network.}
\label{fig:radius}
\end{center}
\end{figure}

We further investigate the distribution of the effective radius in the Steam Community.
The \emph{effective radius} for a node $v$ is defined as the $90$th-percentile of all shortest distances from $v$~\cite{kang11hadi}.
Figure~\ref{fig:radius} shows the frequency distribution of the various radii exhibited in the Steam Community and C-C networks.
The networks have a bi-modal structure with respect to the radius distribution.
The first maximum reflects the nodes that belong to the disconnected components of each network (outsiders~\cite{kang11hadi}).
Nodes in the first dip and under the second maximum belong to the ``core'' of the giant connected component (GCC) of each network.
Nodes on the second maximum are the vast majority of well-connected nodes in the GCC.
Furthermore, the ``whiskers''~\cite{leskovec08community} are the nodes connected to the GCC with longer paths than other nodes and can be seen on the long tail of each plot, thus responsible for its second negative slope. 
Our results on the Steam Community networks verify the observations for the bi-modal behavior of social networks, and in particular LinkedIn~\cite{kang11hadi}.

\subsection{Who Plays with Cheaters?}\label{in-game}

\begin{figure}[htbp]
\centering
	\subfigure[]{
		\includegraphics[width=1.5in]{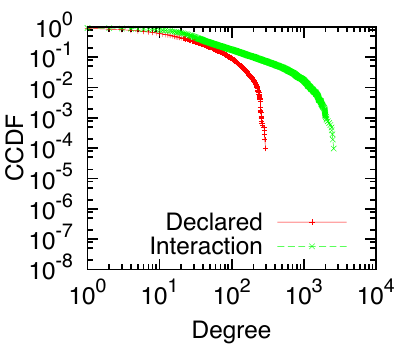}
    		\label{fig:declared_vs_interaction_degree_ccdf}
		}
	\subfigure[]{
		\includegraphics[width=1.5in]{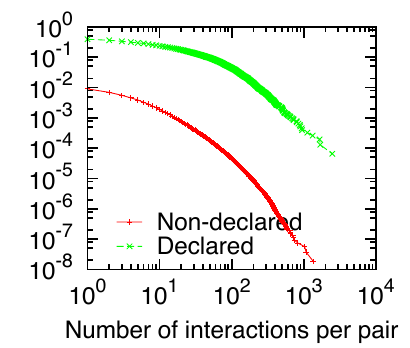}	
		\label{fig:declared_vs_non-declared_interactions_ccdf}
		}
\caption{
(a) CCDF of the declared friendship degree and interaction degree of users from a popular TF2 server.
(b) CCDF of the number of interactions between declared and non-declared pairs of friend players on a popular TF2 server.}
\end{figure}

To investigate if the declared friendships reflect in-game interactions and if cheaters have similar playing patterns with non-cheaters, we studied the 2-month interaction network generated from the TF2 server logs.
Figure~\ref{fig:declared_vs_interaction_degree_ccdf} plots a CCDF of the declared friendship degree as well as the interaction degree for players appearing in the interaction network.
We first note that even on a single server for a single game, players generally interact with considerably more players than they have declared friendships with.
We note, however, that the correlation between the number of declared friends and the number of unique interaction partners is low (Pearson coefficient $0.16$).
This suggests that being popular in the social network does not necessarily translate to an increase in unique interaction partners.

We also compare the number of interactions between declared friends and players that are not declared friends (Figure~\ref{fig:declared_vs_non-declared_interactions_ccdf}).
The plot suggests that players with a declared friendship interact with each other more often than players without a declared friendship.
This indicates that Steam Community friendships are representative of in-game interactions: Steam Community friends are more likely to interact in-game than players who are not friends in the social network.

\begin{figure}[htbp]
\centering
	\subfigure[]{
		\includegraphics[width=1.5in]{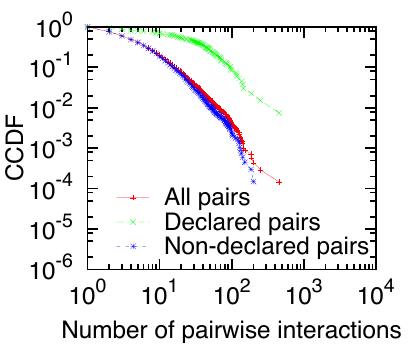}
    		\label{fig:cheater_interactions_ccdf}
		}
	\subfigure[]{
		\includegraphics[width=1.5in]{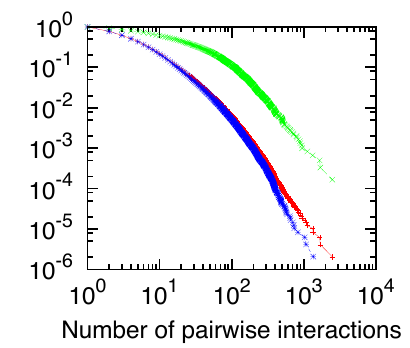}
		\label{fig:non-cheater_interactions_ccdf}
		}
\caption{CCDF of the number interactions between declared and non-declared pairs of players, with at least one user in the pair being a cheater in (a) or a non-cheater in (b), on a popular TF2 server.}
\end{figure}

We next ask: ``Are cheaters ostracized by non-cheaters during gameplay?''
We begin to answer this question by examining how cheaters interact with declared and non-declared friends.
Figures~\ref{fig:cheater_interactions_ccdf} and~\ref{fig:non-cheater_interactions_ccdf} plot the CCDF of the number of pairwise interactions, with at least one member of the pair being a cheater (or non-cheater), for both declared and non-declared Steam Community friendships.

There are two important observations that result from these plots.
First, we see that cheaters, like the population of the server as a whole, are likely to have more interactions with declared friends (than with other players).
Second, 
we notice that interacting pairs with at least one cheater in the pair have fewer absolute interactions than the server as whole.

\begin{figure}[htbp]
\centering
	\subfigure[Gameplay Friends]{
		\includegraphics[scale=0.9]{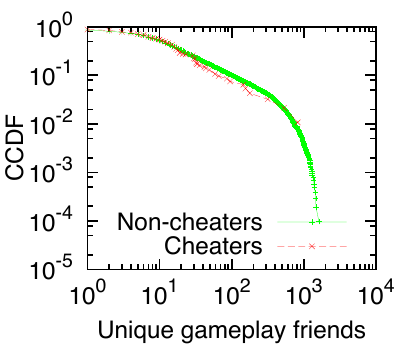}
		\label{fig:cheater_vs_noncheater_friend_degree_ccdf}
	}
	\subfigure[Gameplay Foes]{
		\includegraphics[scale=0.9]{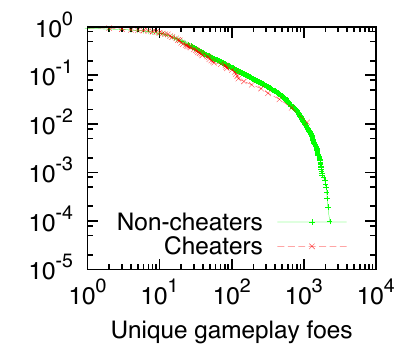}
		\label{fig:cheater_vs_noncheater_foe_degree_ccdf}
	}
\caption{CCDF of the number unique gameplay friends and foes of cheaters and non-cheaters.}
\label{fig:cheater_vs_noncheater_friend_foe_degree_ccdf}
\end{figure}

Since TF2 games are between competing teams, players on the same team have the ability to have cooperative interactions, and those on opposing teams have the opportunity to have antagonistic interactions.
If players had overwhelmingly negative feelings towards cheaters, one would expect cheaters to be involved in fewer cooperative interactions than antagonistic interactions, e.g,. players might banish those with a cheating flag in a form of vigilante justice.
This does not seem to be the case as Figures~\ref{fig:cheater_vs_noncheater_friend_degree_ccdf} and~\ref{fig:cheater_vs_noncheater_foe_degree_ccdf} demonstrate, which present the CCDF of number of unique ``friend''(cooperative) and ``foe''(antagonistic) partners for cheaters and non-cheaters, respectively.
While cheaters tend to have slightly fewer unique gameplay partners than non-cheaters, the difference is negligible, indicating that the cheater/non-cheater status of a player does not hold much weight during active gaming sessions.

\subsection{Are Cheaters in Disgrace?}
\label{sec:cheater-changes}
While aggregate-level information shows little differentiation between cheaters and non-cheaters, the effect of the VAC-ban mark can better be understood by analyzing the transition of profiles from non-cheater to cheater.
We now proceed to answer the following two questions: 1)~Are cheaters shamed by the mark on their permanent record? and 2)~Does the community shun cheaters once their transgressions are revealed?

Of the new cheaters discovered from our re-crawl, $87\%$ had no change in privacy state, and nearly $10\%$ changed their privacy setting from public to a more restrictive setting.
In comparison, in our control group of re-crawled non-cheaters, privacy settings remained unchanged for over $97\%$ of users, and less than $3\%$ changed to a more restrictive setting.
Cheaters seem to choose higher level of privacy once their sins are laid bare, perhaps in the naive hope that a more restrictive setting will provide a measure of protection from a potentially disapproving community.

\begin{figure}[ht]
\includegraphics[scale=1.0]{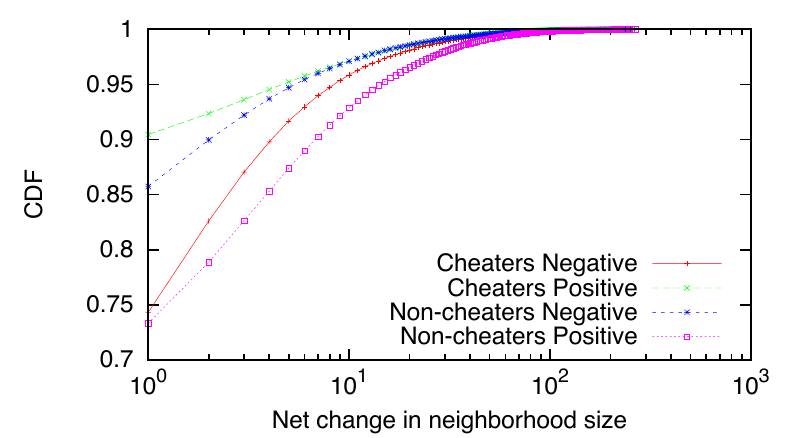}
\caption{CDF of net change in cheaters' and non-cheaters' neighborhood size.}
\label{fig:net_degree_change_cdf}
\end{figure}

But is the local community disapproving? 
Figures~\ref{fig:net_degree_change_cdf} plots the CDF of net change in the degrees for cheaters and non-cheaters.
Of the still public cheaters in our re-crawl, 
$43\%$ had a net loss in degree,
$13\%$ had a net gain in degree, and
$43\%$ had no change in degree.
Of the non-cheaters in our new crawl,
$25\%$ had a net loss in degree,
$36\%$ had a net gain in degree, and
$39\%$ had no change in degree.
While both sets of users exhibited fluctuations in the size of their neighborhoods, more cheaters lost friends than non-cheaters, and more non-cheaters gained friends.
Treated as a whole, cheaters lost nearly twice as many friends as they gained, and non-cheaters gained twice as many friends as they lost.
While non-cheaters continue to gain new friends, cheaters, while not overtly ostracized, appear to be unable to make new friends and may lose a few of their previous ones.

There are several explanations for the changes in neighborhood sizes we observed.
First, evidence suggests that online gamers ``clean up'' their friends lists to remove people they no longer play with~\cite{Xu:2011tu}.
However, because so few users are near the 250 user limit (as seen in Figure~\ref{fig:deg_dist_steam_comm}), we do not believe this is the primary contributing factor to neighborhood size fluctuations.
A second explanation is that the Steam client, by default, issues ``pop up'' notifications that are visible in game whenever a friend starts playing any game.
If you have many friends, the pop ups could become very distracting, possibly prompting users to remove friendships of people they no longer actively play with.
A final explanation, especially with respect to the net loss in cheaters' degrees is that cheaters are deliberately severing their ties once they are caught cheating.
We observed one account in particular that went from 200 to 0 friends after the VAC ban was issued.
``Social suicide'' might account for large decreases in degree, as it is far more probable that the cheater himself deletes friends, rather than each of his friends deleting the cheater.



\subsection{Are Cheaters Close?}\label{geosocial}

In order to quantify the strength of the relationship between cheaters based on the social network and the profile attributes we have, we employ two metrics, one geographical, another social.

\subsubsection{Geographic Closeness}

\begin{figure}[htbp]
\centering
	\includegraphics[scale=1.0]{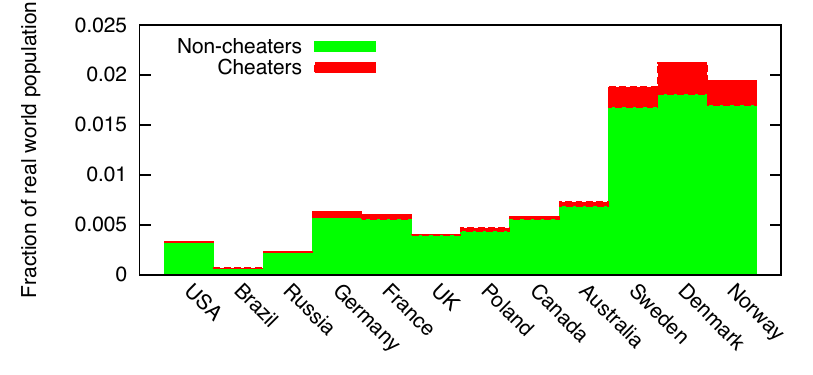}
	\caption{User and cheater populations per country normalized to real world population of said country.
	The countries are arranged along the x-axis in decreasing order of their real-world populations.}
\label{fig:pop_countries_bar_stack}
\end{figure}

The geographical closeness may give quantitative support to the theory proposed in~\cite{Dumitrica:2011jd} according to which our position to cheating is culturally (and thus, to some extend, geographically) based. 
A first observation is that user population on \emph{Steam Community} does not follow real-world geographic population and, more importantly,  cheaters are not uniformly distributed.
Figure~\ref{fig:pop_countries_bar_stack} shows \emph{Steam Community} populations for the twelve countries comprising the top ten user populations and the top ten cheater populations.
The figure shows that cheaters are vastly overrepresented in some locations: for example, there are about 55,000 cheaters in the Nordic  European countries (12.4\% of the playing population of the region), while there are about 39,000 cheaters (3.9\%) in the US.
In particular, we found enough Steam profiles to account for nearly 2.5\% of Denmark's 5.5 million residents, of which cheaters account for nearly 0.5\% of Denmark's population.

\begin{figure}[th]
  \centering
  \includegraphics[scale=1.0]{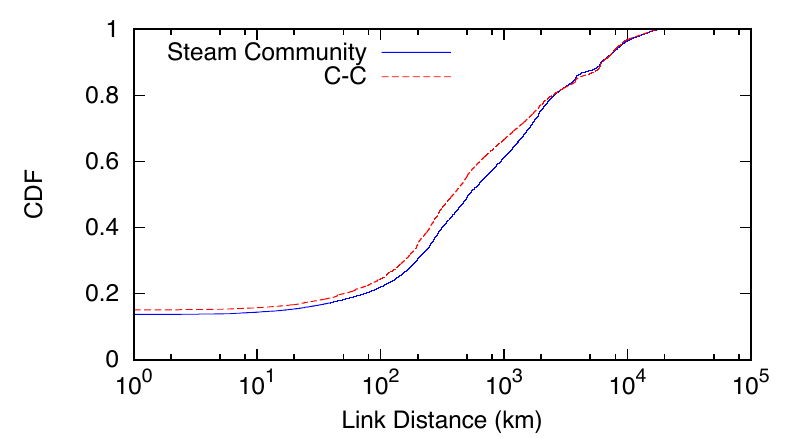}
  \caption{CDF of link length for Steam Community location network.}
  \label{fig:link_length_friendship_cdf}
\end{figure}

We continued our analysis by examining the physical distance between declared friends.
We first constructed the \textit{location network} by including an edge from the social network if and only if both end points had a known location.
This lead to a reduction in the size of the network, which can be seen in Table~\ref{tbl:network_geo_props}, along with the geo-social properties of the resulting location network.
We note that a subgraph composed entirely of cheater-to-cheater relationships has a lower mean distance between nodes and average link length than the graph as a whole.

The cumulative distribution function of link length for the location network is plotted in Figure~\ref{fig:link_length_friendship_cdf}.
About 50\% of all link lengths in all networks are less than 500 km.
In general, cheater-to-cheater relationships tend to be closer than the network as a whole.
This indication that cheaters tend to form relationship with each other over closer distances than the network as a whole, in conjunction with the non-uniform geographic distribution of cheaters leads us to explore how cheaters are positioned in the network from a geo-social perspective

We thus decide to measure node locality, a geo-social metric introduced in~\cite{Scellato:2010tc}.
The node locality of a given node quantifies how close (geographically) it is to \emph{all} of its neighbors in the social graph, and scales it proportional to the geographic network in which the node is embedded.
Thus, a node locality of $1.0$ indicates that a given node is at least as close to all of its neighbors as any other node in the graph is to their neighbors, and a value of $0.0$ indicates that a given node is further away from all its neighbors than any other node in the graph.
Measuring node locality answers two questions for us: 1)~Does the Steam Community exhibit properties of a location-based social network? and 2)~Do cheaters tend to form geographically closer relationships with other cheaters than non-cheaters?

\begin{table*}[htbp]
  \centering
  \begin{tabular}{| l | r | r | r | r | r | r | r |}
    \hline
    Network & N & K & $\langle D_{uv} \rangle$ (km) & $\langle l_{uv} \rangle$ (km) & $\langle NL \rangle$ & $\langle GC \rangle$ \\
    \hline
    Steam Community & 4,342,670 & 26,475,896 & 5,896 & 1,853 & 0.79 & 0.154 \\
    \hline
    Steam Community Cheater-to-Cheater & 190,041 & 353,331 & 4,607 & 1,761 & 0.79 & 0.074 \\
    \hline
    BrightKite & 54,190 & 213,668 & 5,683 & 2,041 & 0.82 & 0.165 \\
    \hline
    FourSquare & 58,424 & 351,216 & 4,312 & 1,296 & 0.85 & 0.237 \\
    \hline
    LiveJournal & 992,886 & 29,645,952 & 6,142 & 2,727 & 0.73/0.71 & 0.146 \\
    \hline
    Twitter & 409,093 & 182,986,353 & 6,087 & 5,117 & 0.57/0.49 & 0.108 \\
    \hline
  \end{tabular}
  \caption{Location network properties: the number of nodes N, edges K, mean distance between users $\langle D_{ij} \rangle$, average link length $\langle l_{ij} \rangle$, average node locality $\langle NL \rangle$, and average geographic clustering coefficient $\langle GC \rangle$.
  The FourSquare, BrightKite, LiveJournal, and Twitter properties are from~\protect\cite{Scellato:2010tc}.}
\label{tbl:network_geo_props}
\end{table*}

\begin{figure}[htbp]
\centering
	\includegraphics[scale=1.0]{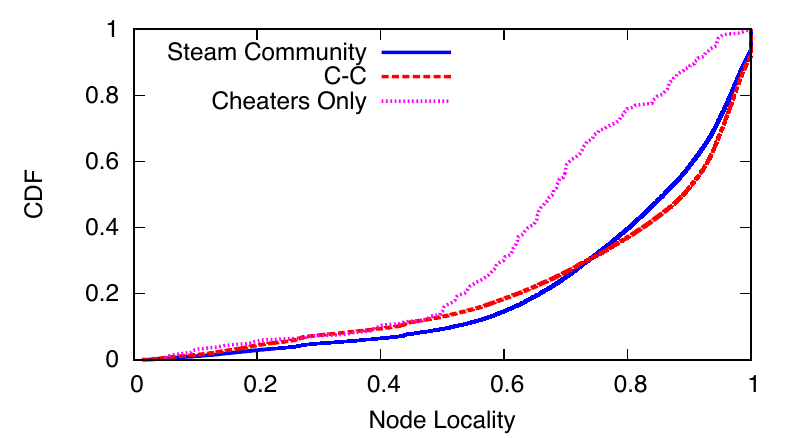}
	\caption{CDF of node locality.}
\label{fig:node_locality_friendship_cdf}
\end{figure}

We plot the CDF of node locality for the location network, the cheater-to-cheater subgraph, as well as just the cheaters within the location network in Figure~\ref{fig:node_locality_friendship_cdf}.
We first note that about 40\% of users in the location network have a node locality of above $0.90$, a phenomena exhibited by other geographic online social networks such as BrightKite and FourSquare~\cite{Scellato:2010tc}.
Next, we observe that while players with low locality are more common in the cheater-to-cheater network than in the whole network, the trend reverses with more players with a node locality of over 0.75 than in the cheater-to-cheater network.
Finally, when considering only the cheaters embedded within the entire network, we see drastically lower node locality, with only about 10\% of cheaters having a node locality greater than $0.90$.

These characterization results lead to three observations: 1)~friendships tend to form between geographically close users; 2)~cheaters tend to form relationships with other nearby cheaters and these links are geographically even shorter than those formed by non-cheaters, and 3)~as evidenced by their lower node locality when considering all friends (cheaters and non-cheaters), cheaters appear to befriend geographically remote non-cheaters. This might indicate that cheaters form relationships with cheaters via a different mechanism than they form relationships with non-cheaters.
While cheater-to-cheater relationships seem to be geographically constrained, their relationships with non-cheaters are over significantly larger distances.

\begin{figure}[th]
 \centering

\includegraphics[scale=1.0]{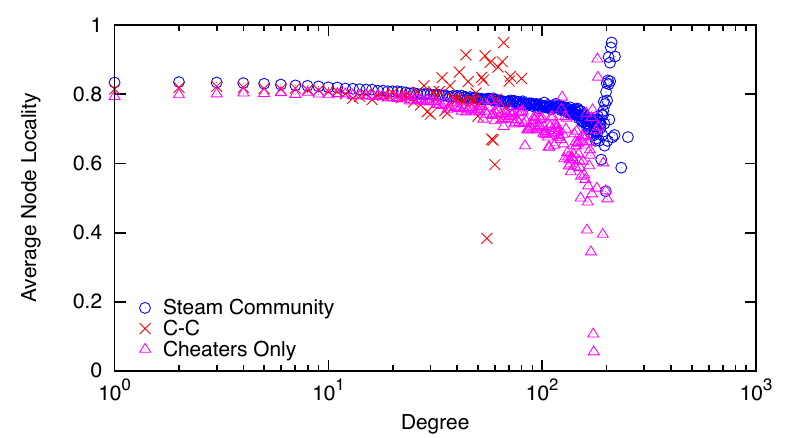}
 \caption{Average node locality as a function of degree.}
 \label{fig:node_locality_by_degree_friendship}
 
\end{figure}

Figure~\ref{fig:node_locality_by_degree_friendship} plots average node locality as a function of degree.
We can see that in all cases, node locality decreases slowly as the degree of the users increases.
This is intuitive as the more friends a user has, the less likely she is to be geographically close to all of them.
However, we note that while there is a decline in locality when the degree of a node gets very close to the limit of 250 friends, the network as a whole has some users of high degree with higher node locality than might otherwise be expected.
This might be indicative of popular users being popular within a geographically constrained portion of the network.
I.e., popular users are popular with people that are geographically close to 
them, but not so much on a global level.

\begin{figure}[th]
 \centering
\includegraphics[scale=1.0]{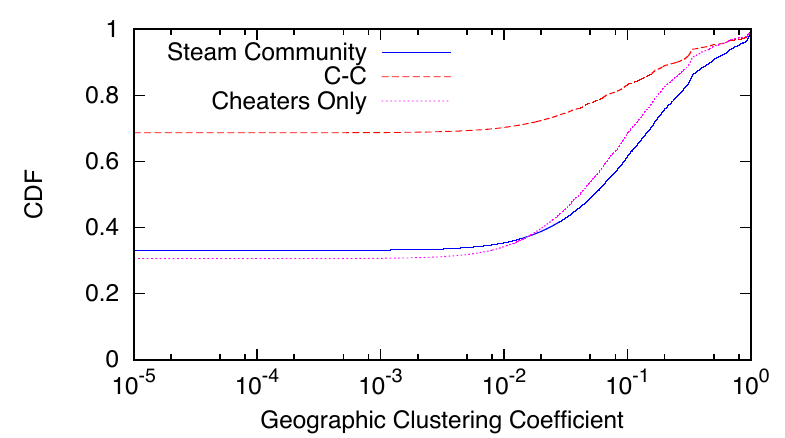}
 \caption{CDF of geographic clustering coefficient.}
 \label{fig:gcc_friendship_cdf}
\end{figure}

The \emph{geographic clustering coefficient} measures how tightly clustered triangles of users are with respect to the geographic distance between connected triples~\cite{Scellato:2010tc}.
The CDF for geographic clustering coefficient is plotted in Figures~\ref{fig:gcc_friendship_cdf}.
Cheaters embedded within both the Steam Community tend to have lower geographic clustering coefficients.
As a whole, around 10\% of the network has a geographic clustering coefficient larger than $0.5$ with 4\% having over $0.9$.
For embedded cheaters, we see only 5\% with a geographic clustering coefficient of over $0.5$ and 2\% greater than $0.9$.
While a larger proportion of cheaters have a geographic clustering coefficient greater than $0.015$ than non-cheaters, this trend quickly reverses, with about 40\% of the whole network having a geographic clustering coefficient greater than $0.1$ versus 30\% of embedded cheaters.
We also see that only about 30\% of users in the cheater-to-cheater network, have a geographic clustering coefficient over $0.01$.
This is contrast to over about 65\% for the entire location network and the cheaters embedded within it.

These numbers are interesting for several reasons.
First, about 30\% of users have a geographic clustering coefficient greater than the average of their respective graphs.
Next, cheaters tend to form more geographically dispersed triples when compared to non-cheaters, drastically so if we consider triples formed exclusively of cheaters.
Finally, while we do see evidence of relatively tight geographic clustering, likely due to latency related quality of service concerns for multi-player gaming, we suspect that relationships are formed around community run servers, thus the distance between triples of users is less important than the distance of triples consisting of two users and a server.
This is a hypothesis we are planning to test in future work.

\begin{figure}[th]
 \centering
\includegraphics[scale=0.9]{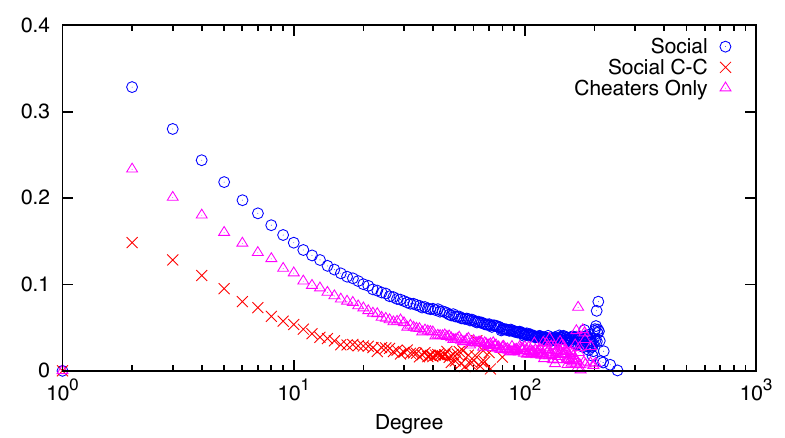}
 \caption{Average geographic clustering coefficient as a function of degree.}
 \label{fig:gcc_by_degree_friendship}
\end{figure}

The geographic clustering coefficient as a function of degree is plotted in Figures~\ref{fig:gcc_by_degree_friendship}.
We can see that the geographic clustering coefficient decreases dramatically as the degree of a node increases.
In other words, the more friends a user has, the less likely those relationships form geographically close triples, an intuitive result.

\subsubsection{Social Closeness}

The second measure of social strength is based on a previous study~\cite{Onnela01052007} that suggests that the overlap between the social neighborhood of two individuals is a good indicator of the strength of their relationship. 
We study the overlap of friends of users in the Steam Community networks to understand whether cheaters exhibit a stronger relationship with other cheaters than fair players do with fair players.
We assess the strength of the relationship between two connected users by the overlap between their sets of friends, computed as follows:
\[
Overlap_{uv} = m_{uv} / ((k_u - 1) + (k_v - 1) - m_{uv})
\]
where $m_{uv}$ is the number of common neighbors between users $u$ and $v$, $k_u$ is the number of neighbors of user $u$ and $k_v$ is the number of neighbors of user $v$.
This overlap is calculated on the friendship network of all users, considering $1.5$ million pairs of cheaters (i.e., all cheater pairs) and $1.5$ million of randomly selected pairs of non-cheaters (i.e., about $2$\% of the existing non-cheater pairs).
We also calculate the overlap of pairs of cheaters (non-cheaters) when considering only their cheater (non-cheater) overlapping contacts instead of their whole neighborhood.

\begin{figure}[htbp]
\begin{center}
	\includegraphics[scale=0.5]{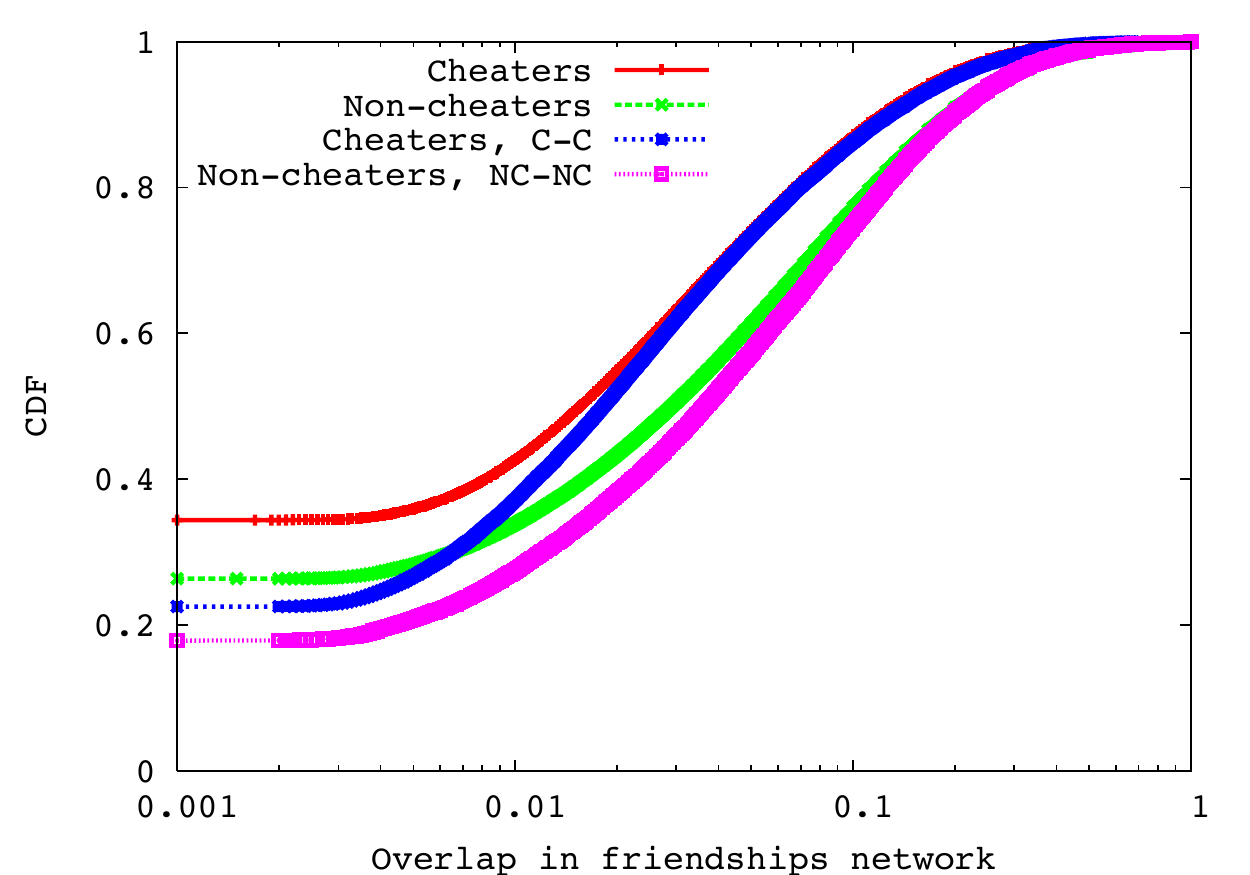}
	\caption{CDF of contacts overlap at the friendship network for cheater and non-cheater pairs.
	We consider all available neighbors and also only the same kind (C-C or NC-NC).}
\label{fig:overlap}
\end{center}
\vspace{-5mm}
\end{figure}

From Figure~\ref{fig:overlap} we observe an increase in the overlap of both cheater pairs in C-C and non-cheater pairs in NC-NC in comparison to the respective overlaps in the overall social network (i.e. considering both types of contacts).
This result demonstrates that relationships are weaker between different types of players, i.e. cheaters with non-cheaters or non-cheaters with cheaters.

\subsection{How do cheaters congregate?}

Steam Community groups allow a set of users to congregate together, providing tools such as event scheduling, group chat, notifications, group level play statistics, and the ability to call out a particular user for exemplary service to the group.
Previous research on player grouping has tended to focus on in-game grouping mechanisms.
For example, guilds in World of War craft have been examined in in~\cite{Ducheneaut06-alonetogether,Ducheneaut:2007if,Ang:2011gv}.
Steam groups differ from most previously studied group constructs for gaming in that they exist separate from any specific game.
Thus, by definition, Steam group relationships persist across games and gaming sessions.
It is also important to note that users are allowed to be a member of more than one Steam group; most in-game grouping constructs enforce either a single group per user rule, with some allowing for hierarchical multi-group organization.

Group membership was determined by querying the crawled users' group memberships.
I.e., the list of groups, as well as each group's members were inferred from information in user profiles.
Because of this there are two caveats: 1)~users with private profiles are not included in the data set and 2)~there are likely some group members that were not discovered via our crawl.
Both of these mean that our results might differ slightly from ground truth, however, they should be pretty close.

We discovered over 5 million users that were members of at least one of over 1 million groups.
Of cheaters with public profiles, 65.4\% are a member of at least one group, vs 58.2\% of non-cheaters.
Table~\ref{tbl:groups} contains additional details of this data set.

Perhaps coincidentally, Ducheneaut et al. found that 66\% of observed World of Warcraft characters were a member of a guild~\cite{Ducheneaut06-alonetogether}, very similar to the observed percentage of cheaters in a group.
This number trended much higher for characters with higher levels, and the indications were that guilds provided both a measurable advantage to members, and also exerted significant ``social pressure'' on members to increase game play time.
Although Steam groups are not directly part of any game play mechanics that we are aware of, it is likely that they fill a similar role as guilds.
As very, very few cheaters are using cheats they created themselves, we suspect that they use Steam groups as a mechanism to discover like minded individuals and distribute cheating information (i.e., to gain a measurable in-game advantage).

\begin{table}

\begin{center}
\begin{tabular}{| l | r |}
  \hline
  Users in at least one group & $5,302,072$ \\
  \hline
  Cheaters in at least one group & $410,704$ \\
  \hline
  Groups & $1,487,551$ \\
  \hline
  Groups with at least 2 members & $1,160,793$ \\
  \hline
  Average group members & $20.9$ \\
  \hline
  Average groups per user & $3.56$ \\
  \hline
  Maximum groups per user & $2,380$ \\
  \hline
\end{tabular}
\end{center}
\caption{Details for groups data set inferred from the Steam Community profiles data set.}
\label{tbl:groups}
\end{table}

\begin{figure}[th]
 \centering \includegraphics[scale=1.0]{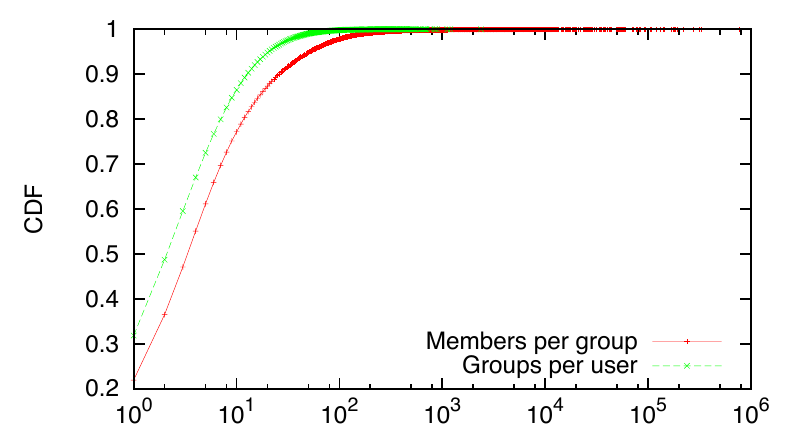}
  \caption{CDF of the members per group and groups per user.}
  \label{fig:users_per_group_groups_per_user_cdf}
\end{figure}

\begin{figure}[th]
 \centering
 \subfigure[]{\includegraphics[scale=1.0]{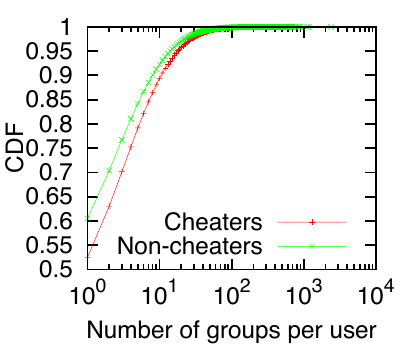}\label{fig:num_groups_per_user_cheater_vs_noncheater_cdf}}
 \subfigure[]{\includegraphics[scale=1.0]{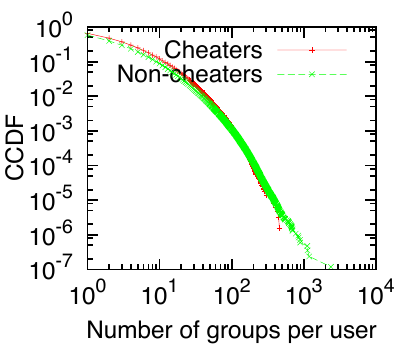}\label{fig:num_groups_per_user_cheater_vs_noncheater_ccdf}}
  \caption{CDF (a) and CCDF (b) of the number of groups per user for cheaters and non-cheaters.}
  \label{fig:num_groups_per_user_cheater_vs_noncheater}
\end{figure}

Figure~\ref{fig:users_per_group_groups_per_user_cdf} plots the members per group and groups per user as a CDF.
90\% of the over 1 million groups had less than 31 members, and only 1\% had more than 150 members.
These membership numbers are similar to the observed sizes of guilds in World of Warcraft where the 90th percentile of the membership distribution was found to be 35~\cite{Ducheneaut06-alonetogether}.
Of the over 5 million users in at least one group, about 35\% were in one, and only one, group with 80\% of users in less than 10 groups.
Surprisingly, cheaters are somewhat more likely to be a member of more groups than non-cheaters, as seen in Figure~\ref{fig:num_groups_per_user_cheater_vs_noncheater}.

\begin{figure}[th]
  \centering
 \includegraphics[scale=1.0]{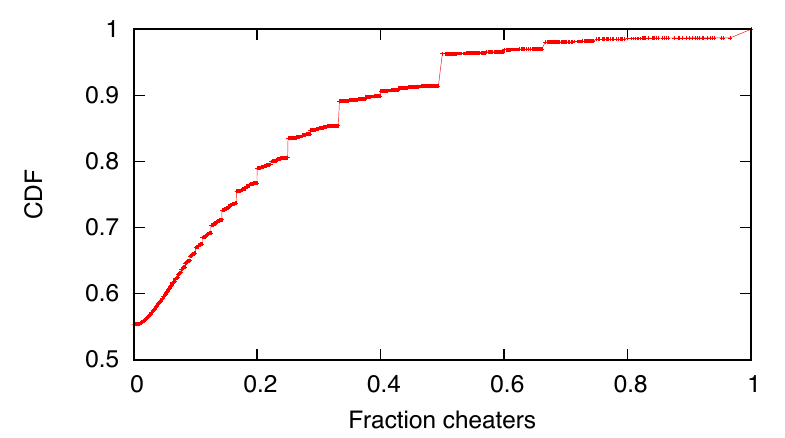}
 \caption{Fraction of group members that are cheaters for groups with a minimum of 2 members.}
 \label{fig:fraction_cheaters_per_group_min_2_members}
\end{figure}

We next examined how cheaters and non-cheaters are represented in the group membership rolls.
We began by discarding groups with less than two members, as at least two members are required for a social relationship to exist.
Of the remaining groups, we plot the fraction of members that are cheaters as a CDF in Figure~\ref{fig:fraction_cheaters_per_group_min_2_members}.
We see that cheaters are not evenly distributed in groups.
In fact, in approximately 10\% of groups non-cheaters are a minority.
We believe these observations indicate that cheaters are making use of the grouping mechanism for different purposes than non-cheaters.

\section{Propagation of Cheating}
\label{propagation}

\begin{figure*}[h!t]
\centering
	\subfigure[CCDF of the number of cheater friends of newly discovered cheaters and a random sample of non-cheaters.]{
		\includegraphics[scale=1.0]{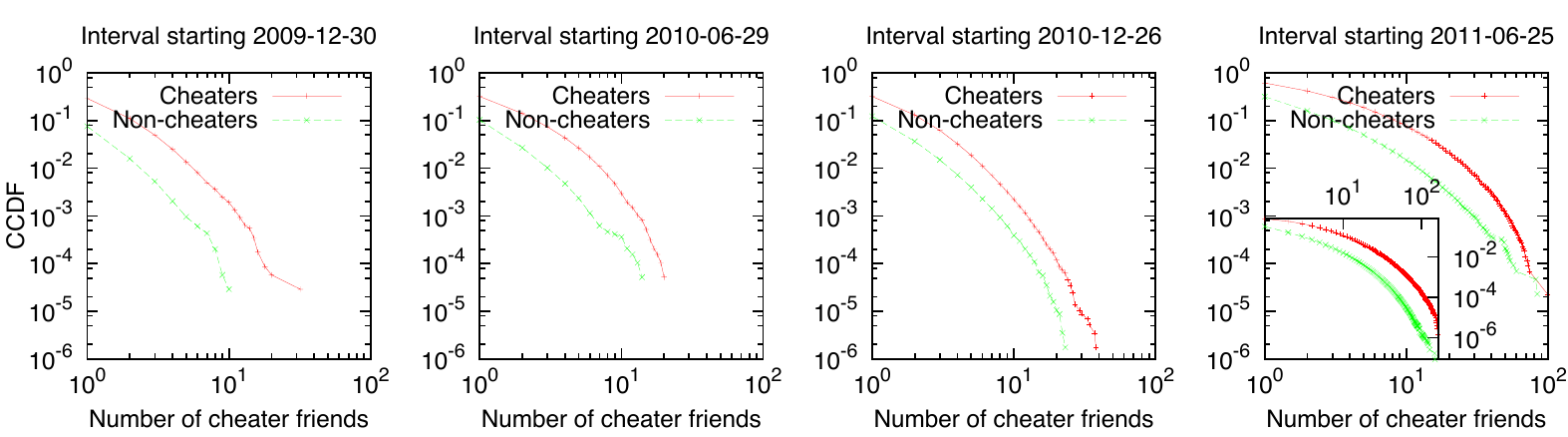}
		\label{fig:integrated_timesliced_c_ccdf}
	}
	\subfigure[CDF of the fraction of cheater friends of newly discovered cheaters and a random sample of non-cheaters.]{
		\includegraphics[scale=1.0]{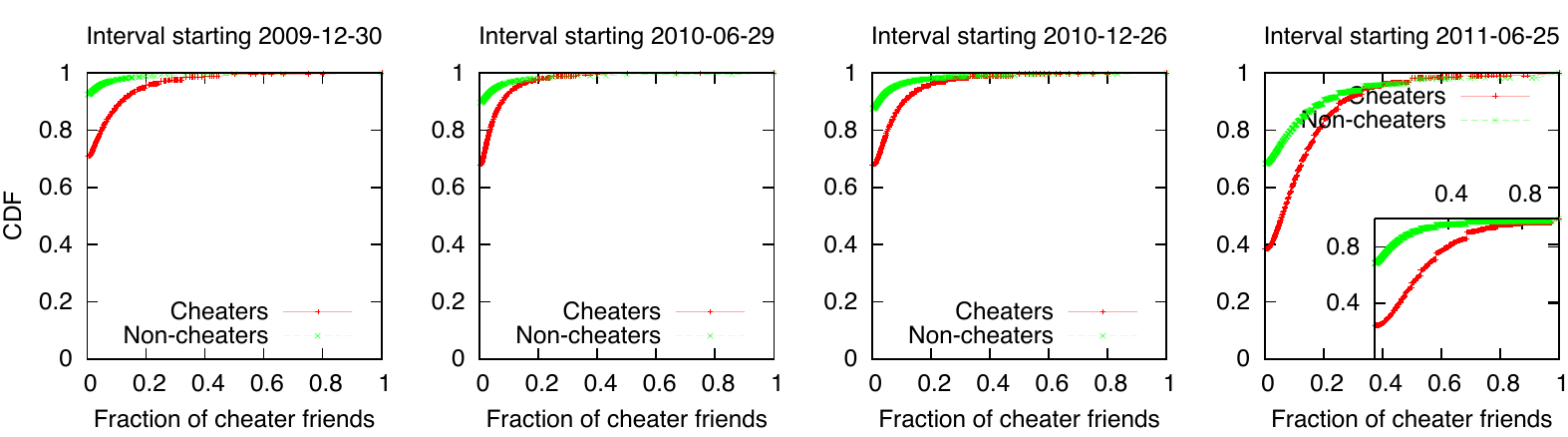}
		\label{fig:integrated_timesliced_fc_cdf}
	}
\caption{The spreading of cheating behavior in the Steam Community over four 180-day time intervals.
The inset plot shows the calculation for the final state of the network, disregarding discovery dates.
A randomly selected control sample of non-cheaters is used for comparison in each time interval.}
\label{fig:cheater_growth_timesliced}
\end{figure*}

How does cheating behavior spread in the Steam Community? 
An approximation of this question may be answered by investigating how cheating bans propagate in the network over time (Section~\ref{time-propagation}). 
Another way to answer this question is to evaluate the position of influence of the cheaters in the network: can they influence indirectly connected players due to their network centrality (Section~\ref{centrality})?

\subsection{How Does Cheating Propagate over Time?}\label{time-propagation}

Based on the observed homophile behavior of cheaters, we hypothesize that the friends of known cheaters are at risk of becoming cheaters themselves.

To test this hypothesis, we explored whether cheaters discovered during a given time interval were more likely to be friends of previously discovered cheaters than to be friends with non-cheaters.
Again, we stress that banned dates must be treated as ``on or before'' as opposed to exact timestamps.
To mitigate the effects of this uncertainty, we chose to examine cheaters discovered over 180-day intervals.

We begin by assuming that all users in the Steam Community friendship network are non-cheaters.
We then initialize the network by marking the $94,522$ users found to have a VAC ban on or before Dec. 29, 2009 (i.e., the earliest date retrieved from vacbanned.com).
For the first time interval between Dec. 30, 2009 and Jun. 28, 2010, $34,681$ players were found to have a VAC ban.
For these users, we calculated and plotted the number and fraction of their cheater friends (i.e., from the $94,522$ cheaters found previously).
We repeat these steps for another 3 time intervals, with $19,294$, $571,975$, and $43,465$ cheaters found in each.
The third interval (starting at Dec. 26, 2010) contains the bulk of cheaters, since their VAC ban was first observed by our initial crawl (and not from vacbanned.com).
However, as shown next, the differentiation between cheaters and non-cheaters holds true for all intervals.
In addition to a best effort approximation of the timestamp of the VAC bans, the data constructed this way has another caveat: the social network is from our March/April 2011 crawl, but we show in Section~\ref{sec:cheater-changes} that the network is quite dynamic. We verified that despite the change in number of friends over time, the trend is preserved: we recalculated the fraction of cheaters in the 1 hop neighborhoods of users based on the state of their relationships
as determined by our October 2011 re-crawl and we found that the non-cheaters CDF to dominate the cheaters CDF as in Figure~\ref{fig:cheater_growth}.

Figures~\ref{fig:integrated_timesliced_c_ccdf} and~\ref{fig:integrated_timesliced_fc_cdf} plot the results of our experiments.
Each subplot represents only the cheaters discovered during the corresponding time interval, and an equal number of randomly sampled non-cheaters.
The number of cheater friends and fraction of cheater friends values are computed based on users that were known to be cheaters \emph{prior} to the start of the interval.
From the plots we see clear evidence that users with both a higher absolute number of cheater friends, as well as those with proportionally more cheater friends are more likely to become cheaters themselves.
In other words, cheating behavior appears to spread through the social network over friendship links.

\subsection{Are Cheaters in Positions of Influence?}
\label{centrality}

Centrality metrics identify important and thus influential nodes in a social network.
To study a player's importance in the Steam Community social network we used two traditional centrality metrics: degree centrality and node betweenness centrality.
The degree centrality of a node is simply the degree of the node in the network, and is thus a local metric.
The betweenness centrality of a node, however, measures the importance of the node in mediating the traffic along the shortest paths between all pairs of nodes.
Betweenness centrality is thus a global graph measure and consequently computationally expensive to calculate, requiring the
calculation of the shortest paths between all pairs of nodes in the network. 
Due to the scale of our graphs, we approximate betweenness centrality using \emph{$\kappa$--path centrality}, a betweenness approximation method proposed in~\cite{alahakoon11sns}.

The $\kappa$--path centrality estimates the betweenness of a node in a network of $n$ nodes and $m$ edges by using random simple walks of length $\kappa$ from random points in the network. 
To reduce the additive error on the betweenness estimation to at most $n^{1/2 + \alpha}$, with probability at least $1-1/n^2$, and in time $O(\kappa^3 n^{2 - 2 \alpha} \log n)$, this process is performed $T = 2\kappa^{2}n^{1-2\alpha}\ln n$ iterations, where $\alpha \in [-1/2, 1/2]$ controls the tradeoff between accuracy and computation time. 
The parameter $\kappa$ was set to $\ln(n+m)$ and $\alpha$ was set to $0.2$, as they offer near optimal performance~\cite{alahakoon11sns}.
To further reduce the running time, we ran the randomized algorithm for computing $\kappa$--path centrality in parallel (the algorithm allows for independent parallelism).  

When we consider all users in the social network, we observe a very high correlation of $0.9731$ between degree and betweenness centrality scores of users.
This high correlation remains consistent when we differentiate on the type of the user.
Thus, when considering cheaters only, the correlation is $0.9817$, and when considering non-cheaters only it is $0.9726$.
This level of correlation implies that if a player has many friends in the Steam Community network, i.e., high degree centrality, 
not only can she influence many other players directly (or locally), but also mediate the information flow between remote players due to her expected high betweenness centrality.
If this player is a cheater, she could facilitate the propagation of cheat code and other deviant behavior to distant parts of the social network.

We extend this analysis and focus only on the most central players in the network and study how many of them are cheaters.
The results, shown in Table~\ref{tab:topN}, demonstrate that the cheaters are under-represented among the most central users of the social network (despite the fact that they have about the same degree distribution as the fair players, as shown in
Figure~\ref{fig:deg_dist}~(a)).
Over $7\%$ of the entire player population in our dataset are cheaters, but they make up less than $7\%$ of the $top$-$1\%$ most central players, and are not adequately represented until we consider the $top$-$5\%$ to $top$-$10\%$ most central players.
Earlier results from Section~\ref{sec:cheater-changes} might provide an explanation for this.
There seems to be social mechanisms that retard the growth of cheaters' social neighborhoods which could be preventing them from entering the $top$-$1\%$ central players in the social network.

\begin{table}
\begin{center}
\begin{tabular}{|c|c|c|}\hline
Top-N\% 		&	DC	&	BC	\\ \hline \hline
0.1			&	3.25	&	5.16\\ \hline
0.5			&	4.46	&	5.95\\ \hline
1.0			&	5.11	&	6.35\\ \hline
5.0			&	7.06	&	7.86\\ \hline
10.0			&	8.20	&	8.58\\ \hline
\end{tabular}
\end{center}
\caption{Percentage of cheaters found in top-N\% of high degree centrality (DC) and betweenness centrality (BC) users in the Steam Community.}
\label{tab:topN}
\end{table}

\section{Summary and Discussion}
\label{sec:discussions}

Online gaming has recently become the largest revenue-generating segment of the entertainment industry, with millions of geographically dispersed players engaging each other within the confines of virtual worlds.
An ethical system is created along with the rules that govern the games. 
Just like in the real world, some players make the decision to circumvent the established rules to gain an unfair advantage, a practice actively discouraged by the industry and frowned upon by gamers themselves.
This paper examined characteristics of these unethical actors in a large online gaming social network.

Due to the scale of our dataset, the majority of our computations used the MapReduce framework via the python mrjob interface for Hadoop on Amazon Elastic MapReduce. 
Our MapReduce stages involved graph pre-processing, gameplay statistics computations, geographical data processing, computing degree distribution, computing intersections of sets,
and computing geo-social metrics. 
Our MapReduce solutions included several MapReduce pipelines (chains of map tasks and reduce tasks) of smaller subtasks. 

At a high level, viewed from the perspective of global network metrics, cheaters are well embedded in the social network, largely indistinguishable from fair players.
This is not entirely unexpected.
Cheaters are still gamers, and even though they are permanently marked, they are still members of the community.
We observed evidence of this by examining both the social network and interaction logs from a multiplayer gaming server, where cheaters were not targeted or treated overly different from non-cheaters.

However, when we examine the transition from non-cheater to cheater, we observe the effects of the cheating brand.
First, cheating behavior appears to spread through a social mechanism, where the presence and the number of cheater friends of a fair player is correlated with the likelihood of her becoming a cheater in the future. 
Consequently, cheaters end up having more cheater friends than the non-cheaters have.
Second, we observed that cheaters are likely to switch to more restrictive privacy settings once they are caught, a sign that they might be uncomfortable with the VAC ban. 
Finally, we found that cheaters lose friends over time compared to non-cheaters, an indication that there is a social penalty involved with cheating.

Cheater distribution does not follow geographical, real-world population density. 
The fact that some regions have higher percentages of cheaters to the player population may suggest that cheating behavior is inspired by the tighter geo-social clustering specific to a geo-social culture. 
Such cheating-prone communities can be the target of more scrutiny or are the result of higher tolerance to cheating
behavior, both in the legislature and in the gaming population.

Our study has consequences for gaming in particular, but also for other online social networks with unethical members. 
In the case of gaming, individual servers can evaluate the cheating risk of a new player by looking at a combination of attributes inferred from the player's profile that include the fraction of VAC-banned friends. 
Our preliminary investigations in this direction show that traditional machine learning algorithms (such as logistical regression, naive Bayes, and decision trees) can classify players as cheaters or non-cheaters with accuracy between $65\%$ and $74\%$. 
More work in this direction is left for the future.

In the case of general online social networks, the findings of our study can be used to better understand the effects of countermeasures to deal with anti-social behavior.
For example, the profiles of users who abuse the available communication tools for political activism or personal marketing, or who appear to automate their actions could be publicly tagged.
Our study gives a preliminary indication that, over time, the reaction of fair users to such information will make it harder to benefit from forms of
anti-social behaviors that attempt to harness network effects.
The fair users tend to have a vested interest in maintaining the quality of the shared social space and will limit the connectivity of the abusing profiles.

\subsection*{Acknowledgements}
We thank 
Yazan Boshmaf, Flavio Vinicius Diniz de Figueiredo, Andreas
Sotirkopoulos, Samer Al-Kiswany, and Chen Shawn of University of
British Columbia for their valuable feedback on an earlier version of
this work.  This research was supported by NSF under Grants No. CNS 
0952420 and CNS 0831785.
 
  \bibliographystyle{abbrv}
  \bibliography{refs}

\end{document}